\documentclass[fleqn,usenatbib]{mnras}

\usepackage{newtxtext,newtxmath}

\usepackage[T1]{fontenc}
\usepackage{ae,aecompl}

\usepackage{graphicx}	
\usepackage{amsmath}	
\usepackage{amssymb}	

\title[Adaptive Optics for Stability in PRV]{Application of Adaptive Optics for Illumination Stability in Precision Radial Velocity Measurements in Astronomical Spectroscopy}

\author[A. J. T. S. Mello et al.]{
Alexandre J. T. S. Mello,$^{1,2}$\thanks{E-mail: ajmello@utfpr.edu.br}
Antonin H. Bouchez,$^{1}$
Andrew Szentgyorgyi,$^{3}$
\newauthor Marcos A. van Dam,$^{4}$
and Henrique Lupinari$^{5}$
\\
$^{1}$GMTO Corporation, 465 N Halstead St, Pasadena, CA 91125, USA\\
$^{2}$Department of Electrical Engineering, Federal University of Technology-Parana, Av. Sete de Setembro, 3165, Curitiba-PR, Brazil\\
$^{3}$Harvard-Smithsonian Center for Astrophysics, 60 Garden St., Cambridge, MA, 02140, USA\\
$^{4}$Flat Wavefronts, 21 Lascelles St., Christchurch 8022, New Zealand\\
$^{5}$University of Sao Paulo, Institute of Astronomy, Geophysics and Atmospheric Sciences, Rua do Matao, 1226, Sao Paulo-SP, Brazil
}

\date{Accepted 2018 September 14. Received 2018 September 11; in original form 2018 May 15}

\pubyear{2018}

\begin{document}
\label{firstpage}
\pagerange{\pageref{firstpage}--\pageref{lastpage}}
\maketitle

\begin{abstract}
Adaptive optics (AO) have been used to correct wavefronts to achieve diffraction limited point spread functions in a broad range of optical applications, prominently ground-based astronomical telescopes operating in near infra-red. While most AO systems cannot provide diffraction-limited performance in the optical passband (400~nm~-- 900~nm), AO can improve image concentration, as well as both near and far field image stability, within an AO-fed spectrograph. Enhanced near and far field stability increase wavelength-scale stability in high dispersion spectrographs. In this work, we describe detailed modelling of the stability improvements achievable on extremely large telescopes. These improvements in performance may enable the mass measurement of Earth Twins by the precision radial velocity method, and the discovery of evidence of exobiotic activity in exoplanet atmospheres with the next generation of extremely large telescopes (ELTs). In this paper, we report on numerical simulations of the impact of AO on the performance of the GMT-Consortium Large Earth Finder (G-CLEF) instrument for the future Giant Magellan Telescope (GMT). The proximate cause of this study is to evaluate what improvements AO offer for exoplanet mass determination by the precision radial velocity (PRV) method and the discovery of biomarkers in exoplanet atmospheres. A modified AO system capable of achieving this improved stability even with changing conditions is proposed.

\end{abstract}




\section{Introduction}

Wavefront control with adaptive optics (AO) is a technique for compensating distortions introduced in the media intervening between a subject being observed and the detectors. The AO technique is applied widely to improve image quality, e.g. in ophthalmology \citep{Marcos2017}. AO systems are prominently used in astronomy to improve image quality at ground based telescopes, where atmospheric phase errors degrade the telescope point spread function (PSF) from a narrow, diffraction-limited Airy pattern (10--20 milliarcsec wide for a 6--10 meter class telescope) into broad seeing disks ($\sim$0.6 arcsec median seeing for a good observing site).

AO systems are used to sharpen imagery \citep{Wizinowich2015}, increase the contrast in coronagraphy \citep{Jovanovic2015} and to reduce the required slit size on spectrographs \citep{Pallavicini2003}. In the latter-most case of spectroscopy, AO may be exploited in different ways to achieve different goals. At a given resolution, implying a certain slit size, AO may simply reduce the seeing disk and put more light into the slit. For fixed spectrograph format, the smaller seeing disk may permit reducing the slit size, thus increasing the achievable resolution of the spectrograph. 

\citet{Rodler2014} have shown that resolution well above 100,000 and ELT scale apertures improve the detectability of biomarkers (especially O2) in exoplanet atmospheres. At a given spectral resolution, reduction in the slit size allows a smaller spectrograph design, with smaller optical elements and a lower construction cost. In the coming era of Extremely Large Telescopes (ELTs), the availability of large optical glass substrates for refractive cameras limits the trade space for optical design.

A final area where AO improves observational capability is the improvement in image stability. Image stability on the spectrograph slit or fibre is critical to the stellar precision radial velocity (PRV) measurements used to discover and measure the mass of exoplanets.

The effect of the atmospheric phase error on image quality and stability is highly wavelength dependent. The compensation of these errors is generally implemented with deformable mirrors, where actuators adjust the figure of the mirror to the shape required to correct the phase errors in real time. For a given telescope size and atmospheric condition, the number of actuators determines the shortest operating wavelength at which diffraction-limited performance can be achieved. For this, and pragmatic considerations of cost, most AO systems on the current generation of 4--10 meter telescopes have been designed to operate at near, mid and thermal infra-red (NIR, MIR, TIR) wavelengths. AO systems operating in the optical band have not been explored extensively, with the exception of the VISAO system deployed at the Clay Telescope, Las Campanas, Chile \citep{Kopon2009}.

In this paper we will focus on the improvement in illumination stability that AO can bring. It will be based in the Giant Magellan Telescope \citep{Johns2012}, which is currently in development phase, and its proposed first light optical band echelle spectrograph, the GMT-Consortium Large Earth Finder (G-CLEF) \citep{Szentgyorgyi2016}. G-CLEF is vacuum-enclosed and fibre-fed to enable PRV measurements, especially for the detection and characterization of low-mass exoplanets orbiting solar-type stars. The recently observed exoplanet around Proxima Centauri will be one of its first targets \citep{Anglada-Escude2016}.

G-CLEF is not being designed with AO in mind, but the availability of a deformable secondary mirror in GMT brings an opportunity to increase precision through illumination stability with little additional cost. The modification required to implement this would be to include a dedicated wavefront sensor for this instrument, which is low cost comparing to the price of a deformable mirror.

In Section \ref{Section:Studies} we discuss effects that impact PRV measurements errors, one of this effects is the illumination stability, that will be the focus of this paper. We then discuss the modelling technique and simulation results of a proposed AO system to improve illuminations stability for G-CLEF in Section \ref{Section:Stability}. Results obtained in an optical bench are presented in Section \ref{Section:Bench}. A control method to deal with turbulence strength variability is presented in section \ref{Section:Control}. We conclude with a discussion of the results and areas of future investigation in Section \ref{Section:Conclusion}.

\section{Illumination Stability Effects on PRV measurements}\label{Section:Studies}

Since the discovery of the first exoplanet orbiting a main-sequence star, 51 Peg B, in 1995 \citep{Mayor1995}, the precision radial velocity (PRV) method has been a powerful tool for the discovery and characterization of exoplanets, especially through mass measurements and determination of orbital elements. These discoveries have been enabled by improvements in wavelength scale calibration with hollow cathode thorium-argon lamps \citep{Kerber2008} or iodine vapour cells \citep{Kambe2002}. The spectrographs themselves are generally fibre-optic fed to isolate and thus stabilize the spectrograph. Early exoplanet discoveries were made with $\sim$10$~\rmn{m~s^{-1}}$ RV precision. As techniques have improved, the attainable precision has increased. The smallest reflex motion measured with a PRV spectrograph is for HD20794c with a K = 0.56$~\rmn{m~s^{-1}}$ \citep{Pepe2011}. HD20794c is an $m~sin(i)= 2.4~\rmn{M\earth}$ planet, orbiting a GV8 with an 18 days period, where i is the unmeasured angle of inclination of the orbit. This level of velocity precision requires sensitivity to Doppler shifts of 1/1000th of a detector pixel, where the detector is generally a charged coupled device (CCD) and the square pixels are 9 or 15~m on a side.

A high priority science goal for the ELTs is the discovery of an Earth twin, i.e. a 1 $\rmn{M\earth}$ planet orbiting a Solar-type star in the habitable zone. An Earth twin would induce a 10$~\rmn{cm~s^{-1}}$ reflex motion in a 1 AU orbit around the Sun, a GV2 star. A factor 5 improvement in the best currently achieved precision is needed to realize this scientific objective.

There are numerous contributors to a PRV error budget \citep{ Podgorski2014}, however a particularly pernicious effect that erodes precision results from either guiding errors or uncompensated atmospheric tip-tilt error. The rays that constitute the seeing disc can be considered to have a unimodal (Gaussian or Moffat) intensity profile. When the intensity profile is well-centred on a slit or optical fibre, the average angle of incidence is zero with respect to the input optical axis of the fibre. However when the centroid of the seeing disk is misaligned with the optical fibre or slit, rays from a particular side of the entrance cone angle defined by the focal ratio of the telescope enter preferentially, and the average angle of entrance will be some finite value larger than zero. While optical fibres are almost perfect scrambler of information concerning the azimuth of entrance into the fibre, there is a surprising persistence of memory about the angle of incidence. A fibre illuminated by an off axis laser produces a nearly perfect annulus in the far field of the output. 

\begin{figure}
  \includegraphics[width=\columnwidth]{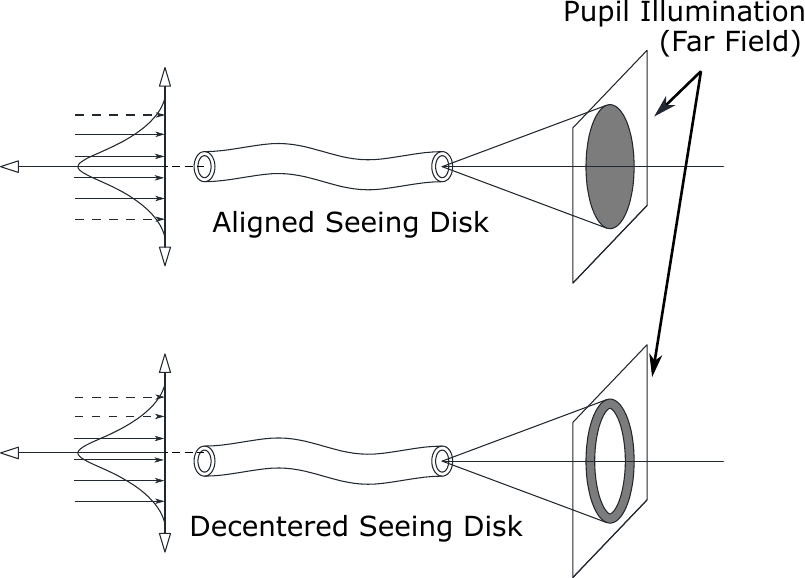}
  \caption{The effect of tip-tilt or misaligned guiding on the pupil illumination in a spectrograph. The intensity distribution of the fibre input, i.e. the seeing disk is indicated. The dashed rays are those that do not enter the optical fibre. The average angle of those rays (not shown for simplicity) determine the light distribution at the fibre exit. Here we indicated, schematically, the effect on the pupil illumination (far field). The fibre end illumination (the near field) will also be affected, but is not illustrated here.}
  \label{figure:fiberalignment}
\end{figure}

In Figure \ref{figure:fiberalignment}, we illustrate how a decentered image can affect the far field in a fibre-fed spectrograph. As seeing or poor guiding jostle the centroid of the seeing disk with respect to the fibre input optical axis, the far field will cause the pupil to be illuminated with varying shaped annuli. This will produce uncalibratable and unpredictable aberrations of the spectrum on the spectrograph detector and distortions of the one-dimensional spectra. Stabilizing the seeing disk on a spectrograph slit or optical fibre input will contribute significantly to the RV precision of any given spectrograph if the slit or fibre size is commensurate with the size of the seeing disk. PRV instruments are required to operate at a resolution of R$\sim$100,000 so as to resolve most of the lines of F,G \& K stars. PRV instruments designed for 6--10~m aperture are generally forced to have slits the size of median seeing disks, so these slits that are overfilled in most seeing conditions. The situation worsen for the apertures of ELTs. Any improvement in input beam stability contributes to RV precision.

An issue is that of how much light gets into the slit or fibre. As we show in this paper, while AO do not achieve the same image quality in the visible passband as they do in the NIR, they do, in fact concentrate visible light considerably to at least a wavelength of 400~nm. Concentrating the light, boosts spectrograph efficiency and throughput, a positive outcome for all types of observational astronomy. Although this boost in efficiency is desirable, it is not the focus of this study, which will concentrate in increasing illumination stability. This will reduce the effects of misalignment of the seeing disk into the fibre entrance.

It is to be noted that there is a literature concerning the advantages of AO for feeding single-mode optical fibres \citep{Crepp2016}. The output of single mode fibres is extremely clean and stable - consisting of a single transmitted mode - and thus many problems associated with multi-mode fibres, especially that illustrated in Figure \ref{figure:fiberalignment}, are eliminated. However, single mode fibre core are extremely small diameter, and thus require the high Strehl ratios achieved with AO operated in "standard" modes when seeing disks are reduced to 10~milliarcsec scales. This level of wavefront correction is not possible in the optical passband with existing technology. Single mode fibres will only be usable in infra-red band for the foreseeable future.

The specific instrument we have studies is that of the GMT - Consortium Large Earth Finder (G-CLEF), the first-light instrument for the Giant Magellan Telescope (GMT) when science operation commence at the GMT in 2024. G-CLEF has been described elsewhere \citep{Szentgyorgyi2016}, however the salient properties of G-CLEF are its optical parameters when operated in PRV mode. These are summarized in Table \ref{table:G-CLEFprop}. The 25.4 m diameter of the GMT primary aperture and the need for R > 100,000 for PRV science has driven the design to one where the pupil is sliced into 7 subapertures, so that the GMT is effectively seven 8.4~m aperture telescopes. Even with this reduction, the largest practical slit size is 0.8~arcsec, while the median seeing is 0.6~arcsec. This means that at least half the time, the slit of the spectrograph will be over-filled.

\begin{table}
\centering
\caption{Optical properties of the G-CLEF fibre slit in PRV mode.}
\label{table:G-CLEFprop}
\begin{tabular}{ccccc}
	\hline
	G-CLEF & f/8 Slit & Telescope & Resolution\\ 
    Mode & Diameter & Plate Scale\\
	\hline 
	PRV & 0.8 arcsec & 1 arcsec mm$^{-1}$ & 105,000 \\ 
	\hline 
\end{tabular}
\end{table}

The AO system that is being design for the GMT will operate in several modes, including Natural Guidestar mode, Laser Tomography mode and GLAO \citep{Bouchez2014}. A logical next step would be to do an experiment with a prototype system on an AO capable large telescope (e.g. the Magellan Clay Telescope). However prototyping and field testing would be quite expensive. We felt a more realistic model of the AO system planned for the GMT was need for a more quantitative assessment of the capabilities of the GMT AO for optical-band spectrographs.

\section{Adaptive Optics Stability Simulations}\label{Section:Stability}

In this section we present results of an AO simulation, taking into account expected turbulence profile, the GMT telescope characteristics, the camera proposed to be used in the wavefront sensor as well as expected noise level. The focus on these simulations is to measure stability gain in the fibre entrance using adaptive optics.

All the simulations we discuss in this paper were done with the YAO software package \citep{Rigaut2013}. YAO is an open source adaptive optics Monte-Carlo simulation tool.

The atmospheric conditions being simulated are based on turbulence profiling made on Las Campanas Observatory by Goodwin \citep{Goodwin2016}. The strength of the turbulence is measured in terms of the Fried parameter, $r_0$. The Fried parameter represents the spatial scale over which the wavefront decorrelates by $\pi$ radians. The larger the Fried parameter, the weaker the turbulence. The expected variation in turbulence strength for the site as measured by Goodwin is $r_0$ between 0.075 and 0.2~m at 500~nm. Table \ref{table:atturb} lists other turbulence parameters.

\begin{table*}
\centering
\caption{Parameters used for atmospheric turbulence simulations.}
\label{table:atturb}
\begin{tabular}{cccccccc}
	\hline
	Layer & 1 & 2 & 3 & 4 & 5 & 6 & 7\\ 
    \hline \hline
    Altitude (km) & 0.025 & 0.275 & 0.425 & 1.25 & 4 & 8 & 13\\
	\hline 
	Wind Speed (m/s) & 5.65 & 5.80 & 5.89 & 6.64 & 13.29 & 34.82 & 29.42\\
	\hline
    Wind Direction (Degrees) & 0.78 & 8.26 & 12.48 & 32.50 & 72.10 & 93.20 & 100.05\\
    \hline
     Fraction & 0.44 & 0.30 & 0.23 & 1.22 & 0.79 & 0.24 & 0.26\\
    \hline
\end{tabular}
\end{table*}

For the wavefront sensor the use of the infra-red camera C-RED One \citep{Gach2016} was considered. This camera has a 320x256 pixels detector, which allows the construction of a Shack-Hartmann wavefront sensor with 60x60 subapertures, each with 4x4 pixels. It is important to point out that the wavefront sensor will operate in the infra-red because the hole visible spectrum must be available for the spectrograph. The target magnitudes considered in simulation are the expected magnitude targets for G-CLEF, ranging from 6 to 12 mag stars.

Other parameters used to simulate the wavefront sensor are: wavelength 1000-1350~nm; 3~\={e} readout noise; 1000~fps readout rate; 1.25 excess noise factor; 0.6 quantum efficiency; sky background 15.5$~mag~arcsec^{-2}$ and photon flux at zero magnitude $3.6\times10^{12}$~$photons~s^{-1}~fullaperture^{-1}$ (corresponding to the collection area of the GMT).

The secondary mirror were simulated with disk harmonics \citep{Verrall1998}. The number of disk harmonics used was empirically tested for the best correction. The simulations results are in terms of encircled energy obtained in the 0.866~arcsec stop.

Table \ref{table:EEvsr0} shows the simulation results for encircled energy in the fibre for two $r_0$ values comparing the gain with and without AO. The use of AO with a wavefront sensor in the infra-red to correct visible light is not common because it is not very efficient. Indeed the gains in term of reduction of the seeing disk size, and as a consequence improvement in captured light in the fixed size fibre, are not large, but the focus here is illumination stabilization.

\begin{table*}
\centering
\caption{Encircled energy in a 0.866 arcsec fibre, as a function of wavelength and turbulence strengh $r_0$ for the cases where Adaptive Optics is used or not used.}
\label{table:EEvsr0}
\begin{tabular}{ccccc}
	\hline \hline
	Wavelength & Without AO & With AO & Without AO & With AO\\ 
    (microns) & $r_0$ = 7.5~cm & $r_0$ = 7.5~cm & $r_0$ = 20~cm & $r_0$ = 20~cm\\
	\hline 
	0.440 & 0.235 & 0.497 & 0.760 & 0.902\\ 
    \hline
    0.550 & 0.245 & 0.515 & 0.774 & 0.904\\
    \hline
    0.640 & 0.253 & 0.532 & 0.784 & 0.906\\
    \hline
    0.790 & 0.267 & 0.561 & 0.797 & 0.911\\
	\hline 
\end{tabular}
\end{table*}

\begin{figure}
		\includegraphics[width=\columnwidth]{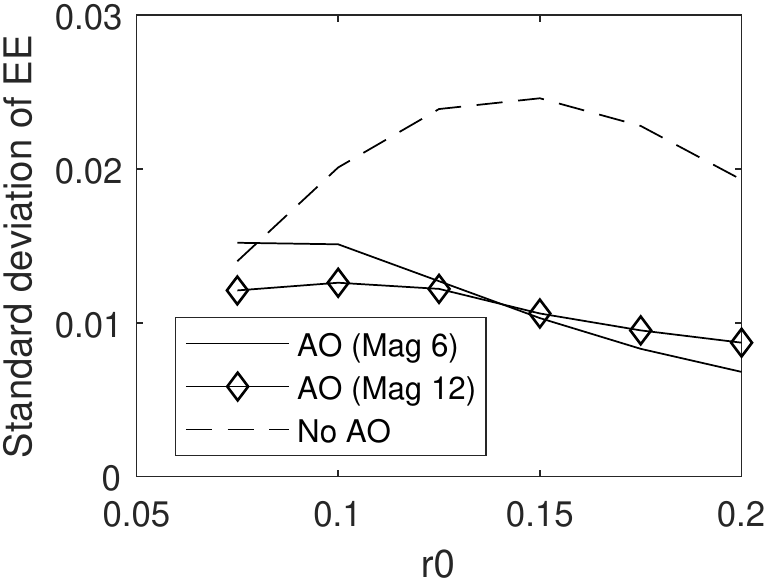}
		\caption{Standard deviation of encircled energy for various turbulence levels for the \textit{B} band (440 nm). The three lines are the cases without the use of AO, and using AO with guidestar magnitudes 6 and 12}
		\label{figure:sd440}
\end{figure}

\begin{figure}
		\includegraphics[width=\columnwidth]{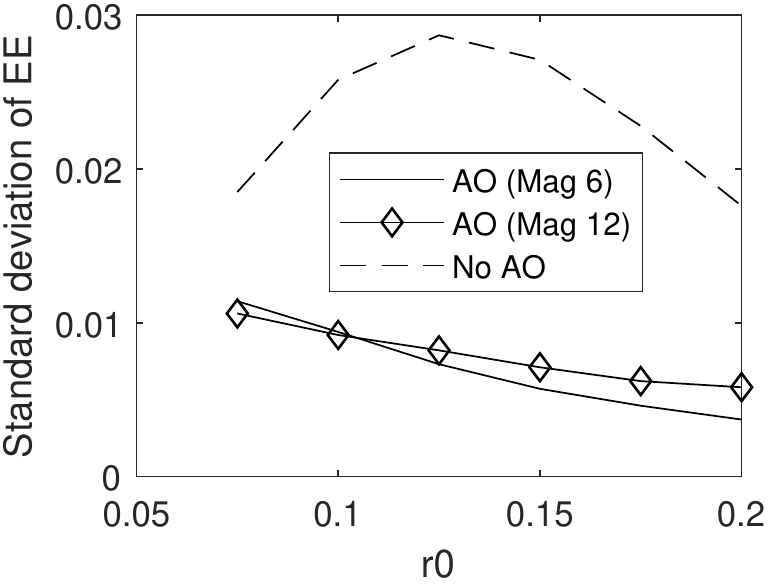}
		\caption{Standard deviation of encircled energy for various turbulence levels for the \textit{I} band (790 nm). The three lines are the cases without the use of AO, and using AO with guidestar magnitudes 6 and 12}
		\label{figure:sd790}
\end{figure}

To determine the stability of the system, the encircled energy variation over time were recorded. Simulations were run with 30~s, simulated time, all using the same simulated turbulence for fairness. The guide stars used in these simulations were of 6 and 12 mag. Results of illuminations stability are show in terms of encircled energy standard deviation for the \textit{B} band (440~nm) in Figure \ref{figure:sd440} and for the \textit{I} band (790~nm) in Figure \ref{figure:sd790}.

The results show that only for very strong turbulence and at short wavelengths AO does not reduce the variability of illumination, but in all other cases the illumination becomes much more stable, making a case for the use of adaptive optics to improve precision.

\section{Optical Bench Results}\label{Section:Bench}

Simulations are not as reliable as experimental results. Some important effects may be overlooked in simulations. For this reason, the experiment done in simulation was repeated in an optical bench, with the objective of determining the gain in illumination stability in a more realistic environment. Although not as complete as the simulation results, the bench results are presented to improve the confidence in the illumination stabilization provided by adaptive optics.

The optical bench experiment was arranged as follows. A collimated laser beam with wavelength $635$~nm passed through a phase-screen, and then was focused on the entrance of an optical fibre. The light exiting the fibre was shone on a power meter.

For the experiment, we used two phase-screens manufactured by Lexitek to simulate moving atmospheric turbulence. One with nominal $0.39$~mm $r_0$ turbulence and the other with $0.39$~mm $r_0$ turbulence after AO correction.

The laser beam had a diameter of $16$~mm. The resulting turbulence would be equivalent of a $20$~cm $r_0$ as viewed by one of the GMT mirrors, with $8.4$~m in diameter. The whole GMT aperture could not be simulated in the optical bench because the size of the phase-screen did not allow it.

\begin{figure}
		\includegraphics[width=\columnwidth]{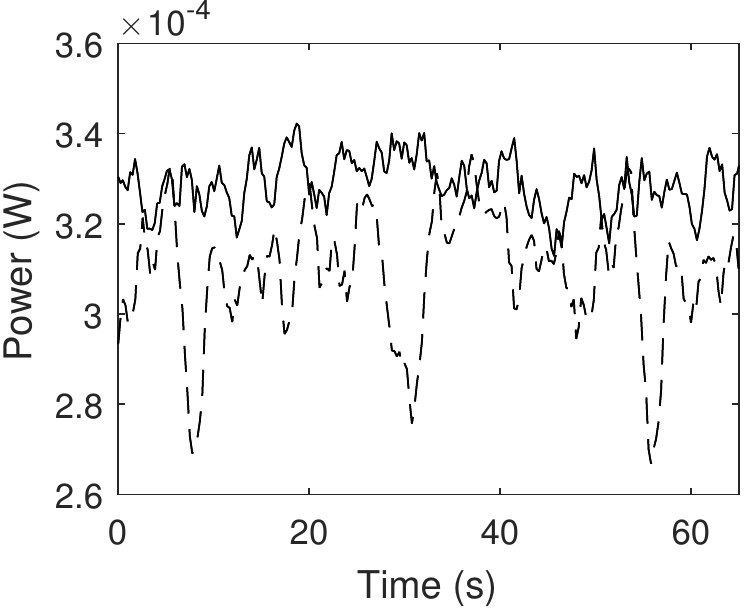}
		\caption{Power Measured on the optical bench experiment. Dashed line is the power over the uncorrected turbulence, straight line is the power over the AO corrected turbulence.}
		\label{figure:benchdata}
\end{figure}

Figure \ref{figure:benchdata} shows the resulting power variations over time. It can be seen that AO correction carries an small improvement in average power, from $3.09\times10^{-4}$~W to $3.29\times10^{-4}$~W. This is a consequence of more light entering the fibre. But the stability is greatly improved, the standard deviation of the power over the uncorrected turbulence is $13.56\times10^{-6}$~W, while for the case with AO corrected turbulence it is $5.70\times10^{-6}$~W.

\section{Stability Control with Set Point}\label{Section:Control}

Sections \ref{Section:Stability} and \ref{Section:Bench} demonstrated how an adaptive optics system improves illumination stability for a fixed $r_0$ value. But what happens when $r_0$ varies? Returning to Table \ref{table:EEvsr0}, it is clear that a variation in $r_0$ will result in a illumination variation, even in a system stabilized by AO.

These variations can be significant on time scales of minutes \citep{Racine1996}, which can impair the illumination stability that we are trying to obtain. In this section, we present a technique to maintain slit illumination constant independent of variations of turbulent strength. 

\begin{figure}
		\includegraphics[width=\columnwidth]{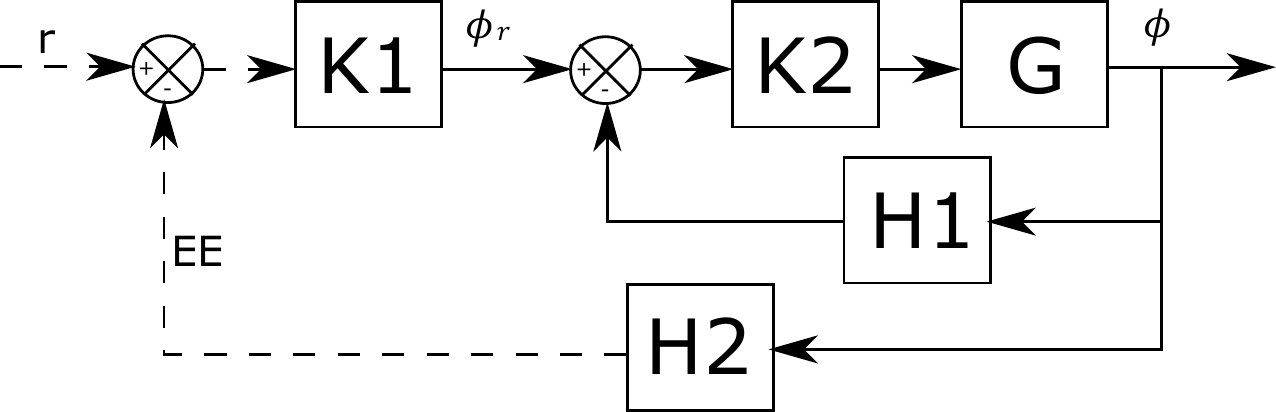}
		\caption{Cascade AO control system for illumination Stability. The telescope system $G$ is controlled by two control loops. $K1$ is the illumination controller, that brings the encircled energy $EE$, measured by $H2$, to the desired level $r$. $\phi_r$ is the wavefront selected by the controller $K1$, which is compared to the wavefront $\phi$ measured by the wavefront sensor $H1$. Finally, $K2$ is a classic AO controller.}
		\label{figure:cascade}
\end{figure}

The proposed technique is an intentional trade-off between encircled energy stability and throughput. We are proposing it because encircled energy stability is more important than absolute throughput for system precision, as absolute throughput can be compensated with a longer exposure. Figure \ref{figure:cascade} shows a diagram of a cascade control system able to compensate for turbulence strength variations.

Dashed signals are scalar values, line signals are vector values. $r$ is the set point value for the encircled energy, which is compared to the measured encircled energy $EE$. The objective of $K1$ controller is to make $EE$ equal $r$. $K1$ is a simple integral controller multiplying a predefined vector. This vector represents a predefined wavefront shape with a high spatial frequency. The integrator will determine the intensity of this wavefront, resulting in a desired wavefront. $G$ represents the telescope system, including the deformable mirror, $H1$ is the wavefront sensor, and $H2$ is a sensor for the encircled energy.

The wavefront sensor $H1$ is implemented as a classic Shack-Hartmann sensor. The EE sensor $H2$ was implemented in the simulator measuring the percentage of light which falls inside the fibres. In the telescope the light in in the guider could be used to obtain this information.

Normally an AO control system is a regulator, that is, it tries to bring the outputs to zero. In this case it is a controller, bringing the output wavefront $\phi$ to $\phi_r$. In an AO regulator, the mirror control voltages vector $\mathbf{v}$ is calculated according to equation~(\ref{e1}), in this case it is modified to the equation~(\ref{e2}). Where $\mathbf{C}$ is the control matrix and $\mathbf{m}$ the wavefront measurements.

\begin{equation}
	\mathbf{v}=\mathbf{C}\times\mathbf{m}
	\label{e1}
\end{equation}

\begin{equation}
	\mathbf{v}=\mathbf{C}\times(\mathbf{m}-\phi_{r})
    \label{e2}
\end{equation}

In effect what this cascade control system does is to modify a classical AO, defined by the inner controller $K2$, which tries to bring the wavefront to a flat form, and converts it to a system that tries to bring the wavefront to a predefined format, with an intensity that will give us the desired encircled energy. The outer controller $K1$, is responsible to define this intensity.

\begin{figure}
		\includegraphics[width=\columnwidth]{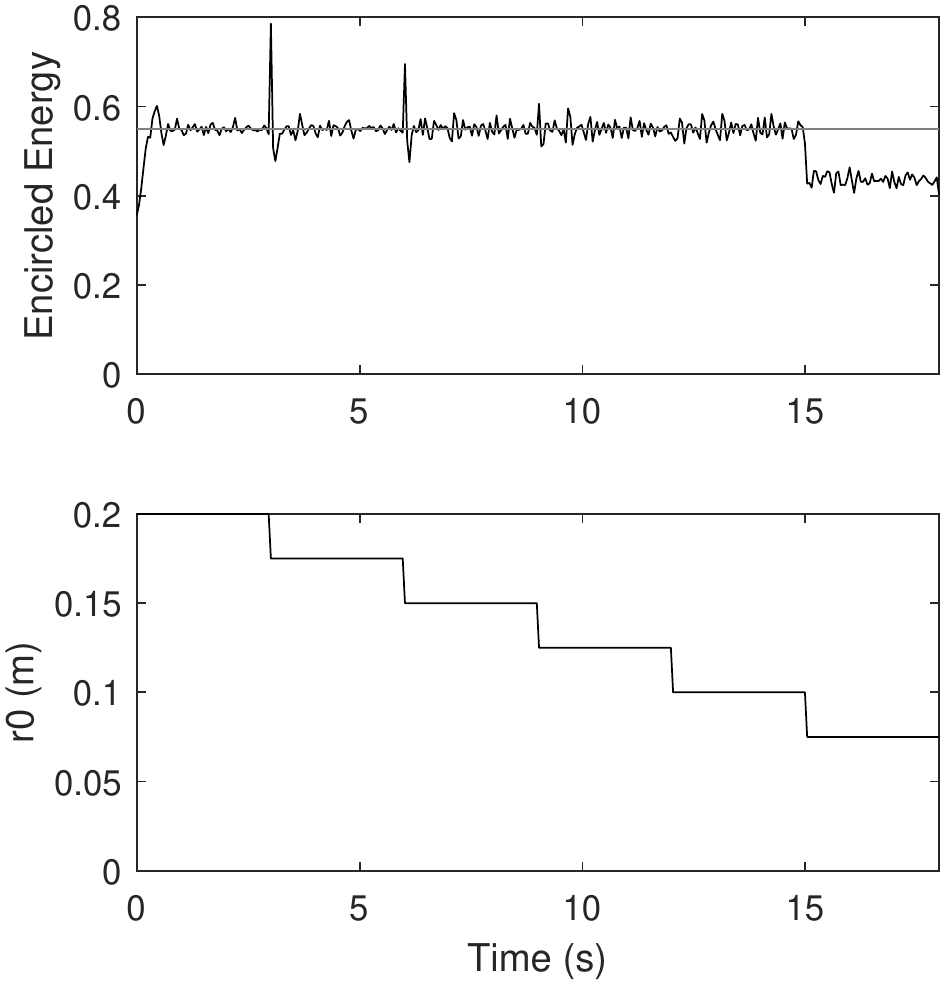}
		\caption{Encircled energy controlled by the cascade control system as $r_{0}$ falls in steps for a magnitude 12 guide star.}
		\label{figure:ctdrop12}
\end{figure}

A simulation of this system was run using a set point of $EE = 0.55$. Figure \ref{figure:ctdrop12} shows results for a 12 magnitude guide star, for the cases where $r_{0}$ is falling in steps.

In this simulation the changes in the turbulence strength were sudden, which resulted in peaks at the change positions because this simple control system can't handle sudden changes. A more refined control could be designed to handle this, but it is not really necessary as this sudden changes doesn't occur in practice, the turbulence evolves slowly over time. In this example the system could not reach the set point only in the case of the stronger turbulence, because the maximum value possible for this adaptive optics system in terms of encircled energy is below the chosen set point. The set-point was chosen to be $EE = 0.55$, but the maximum performance reachable with this $r_{0}$ is $EE = 0.51$. In practice this means the set-point should be chosen according with the stronger expected $r_{0}$, and the system will work as long as the turbulence strength does not surpass this value.

As demonstrated, the proposed system maintains stability even when the turbulence strength changes. In a simple AO implementation the illumination would accompany turbulence strength changes, threatening the desired stability we are trying to accomplish.

\section{Discussion and Conclusions}\label{Section:Conclusion}

In this work we presented a novel idea to use adaptive optics in spectroscopy not aiming on light concentration, but focusing in light stability to increase precision for PRV measurements. Differing from classical AO systems that correct infra-red images using visible light wavefront sensors, we propose to correct visible light using wavefront sensors in the infra-red. 

A study of illumination stability on a realistic simulation environment was presented, and it was shown that the proposed adaptive optics system could be used to improve stability. Only with very strong turbulence and short wavelength (\textit{B} band) there is no improvement in the stability. For most values of turbulence strength and other wavelength bands it was show that the stability is always improved using adaptive optics.

Even with improvements in stability the resulting illumination is still affected by the variations in turbulence strength. To compensate for that a cascade control system to further improve stability over changes in turbulence strength is proposed. The results show that the system would be able to maintain the level of illumination even when turbulence strength changes. This proposed system sacrifices the encircled energy gained with AO, which can be compensated with a longer exposure, to ensure illumination stability, with the aim of increasing precision in PRV measurements.

For future work, it is intended to further improve the control system, and to test it in more realistic conditions. A quantitative study of the precision gained by the increased stability should also be pursued.

\section*{Acknowledgements}

We would like to thank Stephen Eikenberry for support and ideas. This work was supported by the Serrapilheira Institute (grant number Serra-1709-18844). A. Mello thanks the staff at GMTO in Pasadena and IAG, at the University of Sao Paulo, for their hospitality during several visits, where part of this work was developed. H. Lupinari thank FAPESP for the funding of the GMT project and related instrumentation program activities through grant  number 2011/51680-6.




\bibliographystyle{mnras}
\bibliography{references}







\bsp	
\label{lastpage}
\end{document}